\documentclass[twocolumn,aps,prc,floatfix]{revtex4}
\usepackage{graphicx}
\usepackage{dcolumn}
\usepackage{bm}

\begin{document}

\title{Determination of the density region of the symmetry energy probed by the $\pi^-/\pi^+$ ratio}

\author{Gao-Chan Yong$^{1,2}$} \email{yonggaochan@impcas.ac.cn}
\author{Yuan Gao$^{3}$}
\author{Gao-Feng Wei$^{4}$}
\author{Ya-Fei Guo$^{1,2}$}
\author{Wei Zuo$^{1,2}$}

\affiliation{%
$^1${Institute of Modern Physics, Chinese Academy of Sciences, Lanzhou 730000, China} \\
$^2${School of Nuclear Science and Technology, University of Chinese Academy of Sciences, Beijing 100049, China} \\
$^3${School of Information Engineering, Hangzhou Dianzi University, Hangzhou 310018, China} \\
$^4${Shanxi Key Laboratory of Surface Engineering and Remanufacturing, School of Mechanical and Material Engineering, Xi'an University, Xi'an 710065, China}
}%

\begin{abstract}

The nuclear symmetry energy around or below saturation density has been extensively studied and roughly pinned down, while its behavior at suprasaturation densities is rather uncertain. Related experimental studies are being carried out or planned at facilities that offer radioactive beams worldwide. Towards the physical goal of probing the nuclear symmetry energy at high densities, $\pi$ measurements in the medium nuclei $^{132}$Sn+$^{124}$Sn collisions at 300 or 200 MeV/nucleon incident beam energies are ongoing at Radioactive Isotope Beam Facility (RIBF) at RIKEN in Japan. However, our studies show that the observable $\pi^-/\pi^+$ ratio in the $^{132}$Sn+$^{124}$Sn reactions at 300 or 200 MeV/nucleon incident beam energies mainly probes the symmetry energy around the saturation density. Only the $\pi^-/\pi^+$ ratio in the heavy reaction system and at relatively high incident beam energies may mainly probe the symmetry energy at suprasaturation densities.

\end{abstract}

\maketitle

\section{Introduction}

The nuclear symmetry energy, i.e., the single nucleon energy
changes as one replaces protons in nuclear matter
with neutrons \cite{BHF1}, has been extensively studied in nuclear physics community
\cite{Bar05,LCK08,topic} simply because in a density range of 0.1 $\sim$ 10 times
nuclear saturation density, the symmetry energy determines the
birth of neutron stars and supernova neutrinos \cite{Sumiyoshi95},
a range of neutron star properties such as the cooling rates, the
thickness of the crust, the mass-radius relationship, and the
moment of inertia \cite{Sum94,Lat04,Ste05a,Lattimer14}. The
nuclear symmetry energy also plays crucial roles in the evolution
of the core-collapse supernova \cite{Fischer14} and the astrophysical
r-process nucleosynthesis \cite{Nikolov11}.

To constrain the symmetry energy in a broad density region, besides
the studies in astrophysics \cite{apj10,fatt13,fatt14,GW17}, many terrestrial
experiments are being carried out or planned using a wide variety
of advanced new facilities, such as the Facility for Rare Isotope
Beams (FRIB) in the US \cite{frib}, the Facility for
Antiproton and Ion Research (FAIR) at GSI in Germany \cite{fair}, the Radioactive Isotope Beam Facility
(RIBF) at RIKEN in Japan \cite{exp1,exp2}, the Cooling Storage Ring on the Heavy Ion Research Facility at IMP (HIRFL-CSR) in China \cite{csr},
the Korea Rare Isotope Accelerator (KoRIA) in Korea \cite{koria}, etc.

Nowadays the nuclear symmetry energy and its slope around saturation density
have been roughly pinned down
\cite{lihan13,jpj2014}. However, due to complexity of nuclear force, the density-dependent symmetry energy by the power-law fit at lower density and higher density
may require different exponents \cite{zuo2002}. The symmetry energy
at high densities is thus still controversial
\cite{guo2013,guo2014}.
To probe the density-dependent symmetry energy in heavy-ion collisions, many potential
observables have been identified, such as $\pi^-/\pi^+$ ratio \cite{NPDF2,
pion1,pion2,pion3,pion4,pion5,pion6,pion7}, energetic photon and
$\eta$ productions \cite{yong14,yongph}, free neutron to proton ratio
$n/p$ \cite{np1,np2,np3}, $t/^3He$ \cite{t1,t2}, isospin
fractionation \cite{frac1,frac2,frac3,np2}, isospin diffusion \cite{iso04,iso05}
and neutron-proton
differential flow \cite{NPDF1,NPDF3}, etc.
Aiming at probing the symmetry energy at high densities by the $\pi^-/\pi^+$ ratio, Sn+Sn reactions at 300 or 200 MeV/nucleon experiments are being carried out at RIKEN in Japan \cite{exp1,exp2,exp3}.

Does the $\pi^-/\pi^+$ ratio in heavy-ion collisions at intermediate energies always probe the high-density symmetry energy? And in what conditions (reaction system and beam energy) does the $\pi^-/\pi^+$ ratio probe the symmetry energy at low or high densities?
In this study, we try to determine the density region of the symmetry energy probed by the $\pi^-/\pi^+$ ratio in medium and heavy reaction systems. Our study shows that the $\pi^-/\pi^+$ ratio in the medium nucleus-nucleus collisions below 400 MeV/nucleon incident beam energies generally probes the symmetry energy around the saturation density. Only in the heavy reaction system and with relatively high incident beam energies, the $\pi^-/\pi^+$ ratio may be used to probe the symmetry energy at suprasaturation densities.

\section{The methodology and results}

\begin{figure}[th]
\centering
\includegraphics [width=0.5\textwidth]{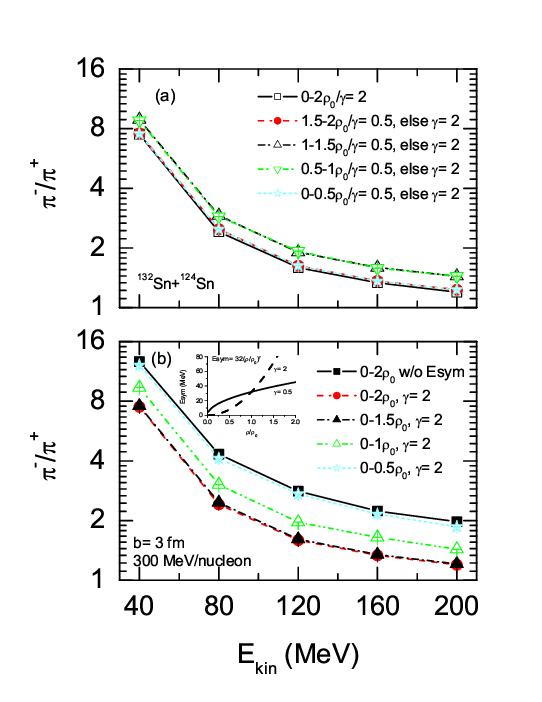}
\caption{\label{rs} (Color online) Relative sensitivity of the
symmetry energy sensitive observable $\pi^-/\pi^+$ ratio as a function
of kinetic energy (in center of mass frame) in the $^{132}$Sn+$^{124}$Sn reaction at the incident beam energy 300 MeV/nucleon. Panel (a) utilizes switching method while Panel (b) with adding method, see text for details. The parameter $\gamma$ = 0.5 and 2 denote soft and stiff symmetry energies, respectively.}
\end{figure}
In this study, pion production in our used transport model is the same as that shown in
Ref. \cite{yong2017} but without medium corrections of both elastic and inelastic baryon-baryon scattering cross sections. We use the Skyrme-type parametrization for the isoscalar mean field \cite{sky2016,bertsch,todd2005}, i.e., $U(\rho)=-0.232(\rho/\rho_{0})+0.179(\rho/\rho_{0})^{1.3}$.
We did not use the Gogny-inspired parameterization for the single particle potential owing to our limited computational power (as we discussed later, this choice would not affect our physical results here). For the purpose of determining in which density region the observables are sensitive to the symmetry energy, a specific density-dependent symmetry energy form $E_{sym}= 32(\rho/\rho_0)^{\gamma}$ is used \cite{fan2018}.
Pion production in heavy-ion collisions at intermediate energies mainly connects with the four channels, i.e., $NN \rightleftharpoons N \Delta$ and $\pi N \rightleftharpoons \Delta$. Although pions are mainly produced at higher densities in heavy-ion collisions, they may suffer from rescatterings and absorption by surrounding nucleons at  high and low densities. Therefore, pions produced in heavy-ion collisions at intermediate energies may be not a good messenger of the high-density symmetry energy of nuclear matter.

One may argue that the energetic pions produced in heavy-ion collisions at intermediate energies should carry the information of high density nuclear matter. However, the energetic particles produced in heavy-ion collisions are in fact pushed by the gradient force of the mean-field potential. The gradient force is strong mainly at the border of the density region while not always in the core of the density region.
From this point, energetic particles may not always carry the information of dense matter at maximum density. Therefore one should make study to determine the density region of the symmetry energy probed by the $\pi^-/\pi^+$ ratio.

In order to know in which density region the $\pi^-/\pi^+$ ratio shows maximum sensitivity to the symmetry energy, one way is similar to the studies in Refs.~\cite{liuy051,liuy052}, i.e., in
the whole density region ($0 < \rho < \rho_{max}$) one chooses a specific form of the density-dependent symmetry energy as the standard calculation.
To get the relative sensitivity of an observable in different
density regions, one changes the density-dependent symmetry energy in different density regions, respectively. Since the value of the symmetry potential is in fact very small compared with the nucleon isoscalar potential, this operation would not cause a ``sudden jump'' of the motion of nucleons in these regions.
Figure~\ref{rs}(a) shows the relative sensitivity of the symmetry energy sensitive observable $\pi^-/\pi^+$ ratio in different density regions.
Since almost all the isospin-dependent transport models show that the kinetic energy distribution of the $\pi^-/\pi^+$ ratio is more or less affected by the symmetry energy, we in this study mainly show the density region for the symmetry energy probed by the kinetic energy distribution of the $\pi^-/\pi^+$ ratio. The relationship between the pion kinetic energy spectra and the symmetry energy has been demonstrated by S$\pi$RIT Collaboration \cite{spi2017}.
We use the symmetry energy parameter $\gamma = 2 $ in the density-dependent symmetry energy form Esym = 32$(\rho/\rho_{0})^{\gamma}$ (Esym is the total symmetry energy, including the kinetic part and the potential part) as the standard calculation and respectively change the parameter $\gamma$ from 2 to 0.5 in different density regions to obtain a series of $\pi^-/\pi^+$ ratios. The choices of the symmetry energy parameters $\gamma$ = 2 and 0.5 are just for convenience of the research.
It is seen that below 0.5$\rho_{0}$ or above 1.5$\rho_{0}$, effects of the symmetry energy on the $\pi^-/\pi^+$ ratio are in fact negligible. The effects of the symmetry energy in density regions of 0.5$\rho_{0}$-$\rho_{0}$ and $\rho_{0}$-1.5$\rho_{0}$ are both obvious and roughly equal. And the corresponding values of the $\pi^-/\pi^+$ ratio are both higher than the standard calculation with $\gamma$ = 2. This is because the change of $\gamma$ from 2 to 0.5 soften the symmetry energy.
The soften symmetry energy causes a high value of the $\pi^-/\pi^+$ ratio \cite{pion4}.
The roughly equal effects of the symmetry energy in the two density regions of 0.5$\rho_{0}$-$\rho_{0}$ and $\rho_{0}$-1.5$\rho_{0}$ reveal that, the
symmetry energy sensitive observable $\pi^-/\pi^+$ ratio in $^{132}$Sn+$^{124}$Sn reaction at 300 MeV/nucleon in fact just probes the density-dependent symmetry energy around saturation density $\rho_{0}$.

To confirm the above physical results, the other method to demonstrate the same question is adding the symmetry energy from zero to maximum density with 0.5$\rho_{0}$ density interval. For each case, because the symmetry energy changes smoothly as the density increases, this method seems more acceptable.
Figure~\ref{rs}(b) shows the relative sensitivity of the $\pi^-/\pi^+$ ratio in different density regions by adding the symmetry energy Esym = 32$(\rho/\rho_{0})^{\gamma= 2}$ from zero to maximum density with 0.5$\rho_{0}$ density interval. As expected, the value of the $\pi^-/\pi^+$ ratio almost does not change when adding the symmetry energy in the density regions 0-0.5$\rho_{0}$ or 1.5$\rho_{0}$-2$\rho_{0}$. While in the density regions of 0.5$\rho_{0}$-$\rho_{0}$ and $\rho_{0}$-1.5$\rho_{0}$, the effects of the symmetry energy are both obvious. And the effects of the symmetry energy below saturation density are comparable to that above saturation density. The nuclear symmetry potential is more attractive for protons than neutrons and causes neutron-deficient matter in heavy-ion collisions, which corresponds a low value of the $\pi^-/\pi^+$ ratio. We thus see the value of the $\pi^-/\pi^+$ ratio generally decreases as the symmetry energy is added with 0.5$\rho_{0}$ density interval.

Further studies show that, in the $^{132}$Sn+$^{124}$Sn reaction the $\pi^-/\pi^+$ ratio mainly probes the high-density symmetry energy only when the incident beam energies are larger than 400 MeV/nucleon. And the effects of the symmetry energy on the $\pi^-/\pi^+$ ratio gradually decrease as the incident beam energies increase. The impact parameter and kinematic cuts of pion emission in the $^{132}$Sn+$^{124}$Sn reactions do not evidently affect the determination of the density region of the symmetry energy by the $\pi^-/\pi^+$ ratio.

\begin{figure}[t]
\centering
\includegraphics [width=0.5\textwidth]{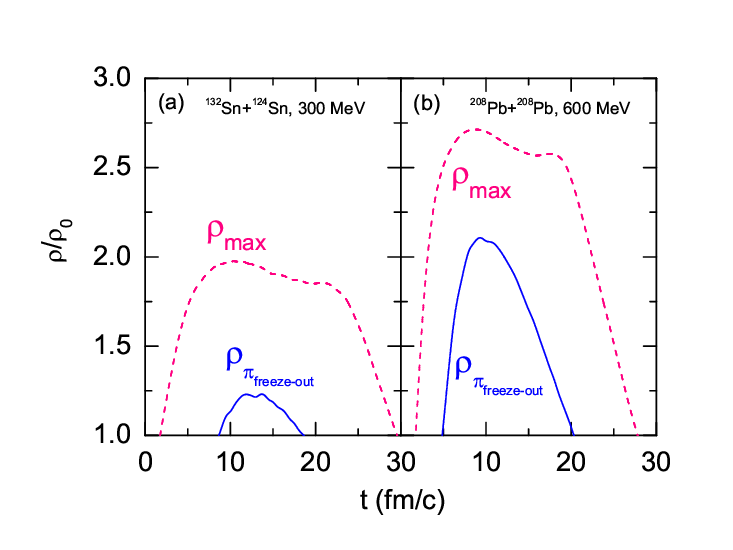}
\caption{\label{pifree} (Color online) Pion freeze-out local baryon density as well as central maximum baryon density as a function of time. Panel (a) shows $^{132}$Sn+$^{124}$Sn reaction at 300 MeV/nucleon with an impact parameter b= 3 fm. Panel (b) denotes $^{208}$Pb+$^{208}$Pb reaction at 600 MeV/nucleon with an impact parameter b= 0 fm.}
\end{figure}
To see why the $\pi^-/\pi^+$ ratio in the $^{132}$Sn+$^{124}$Sn reaction at incident beam energy 300 MeV/nucleon does not probe the symmetry energy at high densities, we plot in Figure~\ref{pifree},  pion freeze-out local baryon density as a function of time.
From Figure~\ref{pifree}(a), it is seen that, for the $^{132}$Sn+$^{124}$Sn reaction at 300 MeV/nucleon, pion freeze-out average local baryon density (the local density where pion does not re-scatter with other particles) is below 1.25$\rho_{0}$ although the maximum baryon density can reach about 2$\rho_{0}$. It means that less pions directly escape the dense matter with density above 1.25$\rho_{0}$ due to the re-scatterings with other nucleons. It also means that below 1.25$\rho_{0}$, produced pions in nuclear matter can easily escape the matter.
So it is not hard to understand why the $\pi^-/\pi^+$ ratio in the $^{132}$Sn+$^{124}$Sn reaction at the incident beam energy 300 MeV/nucleon does not probe the nuclear symmetry energy around twice saturation density.

To probe the high-density symmetry energy by the $\pi^-/\pi^+$ ratio, it seems that one has to use heavier reaction system and with relatively high incident beam energies. Figure~\ref{pifree}(b)
shows the pion freeze-out local baryon density as a function of time in the $^{208}$Pb+$^{208}$Pb reaction at 600 MeV/nucleon. Compared with the light reaction system $^{132}$Sn+$^{124}$Sn at 300 MeV/nucleon, the maximum pion freeze-out local baryon density in the $^{208}$Pb+$^{208}$Pb reaction at 600 MeV/nucleon is above 2$\rho_{0}$, which is much larger than that of the light system. This means that below 2$\rho_{0}$, pions can easily escape the matter.
The $\pi^-/\pi^+$ ratio in the heavy reaction system at high beam energies is thus expected to probe the nuclear symmetry energy at high densities.

\begin{figure}[t]
\centering
\includegraphics [width=0.5\textwidth]{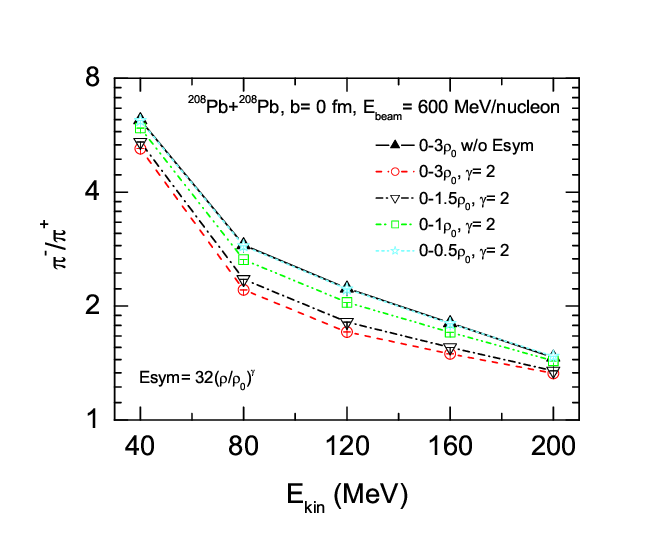}
\caption{\label{pb208} (Color online) Relative sensitivity of the
symmetry energy sensitive observable $\pi^-/\pi^+$ ratio as a function
of kinetic energy (in center of mass frame) in the $^{208}$Pb+$^{208}$Pb reaction at the incident beam energy 600 MeV/nucleon, see text for details.}
\end{figure}
Figure~\ref{pb208} shows the relative sensitivity of the $\pi^-/\pi^+$ ratio to the density-dependent symmetry energy in the $^{208}$Pb+$^{208}$Pb reaction at the incident beam energy 600 MeV/nucleon.
It is seen that even for the $^{208}$Pb+$^{208}$Pb system, the effect of the symmetry energy in density region between 0.5-1.0$\rho_{0}$ is sizable, which is about half of that between 1.0-1.5$\rho_{0}$. Thus even for the heavy system at 600 MeV/nucleon the effects of the symmetry energy at sub-saturation densities play a role.
Figure~\ref{pb208} clearly shows that the effects of the symmetry energy at suprasaturation densities are much larger than that below saturation density. Above 1.5$\rho_{0}$, the symmetry energy still has some effects on the $\pi^-/\pi^+$ ratio. We also simulated the $^{208}$Pb+$^{208}$Pb reaction at 300 MeV/nucleon, and find that the $\pi^-/\pi^+$ ratio mainly probes the symmetry energy around saturation density.
Therefore, to probe the symmetry energy at suprasaturation densities, it seems that one should carry out heavy reaction system experiments and with relatively high incident beam energies.

\begin{figure}[t]
\centering
\includegraphics [width=0.5\textwidth]{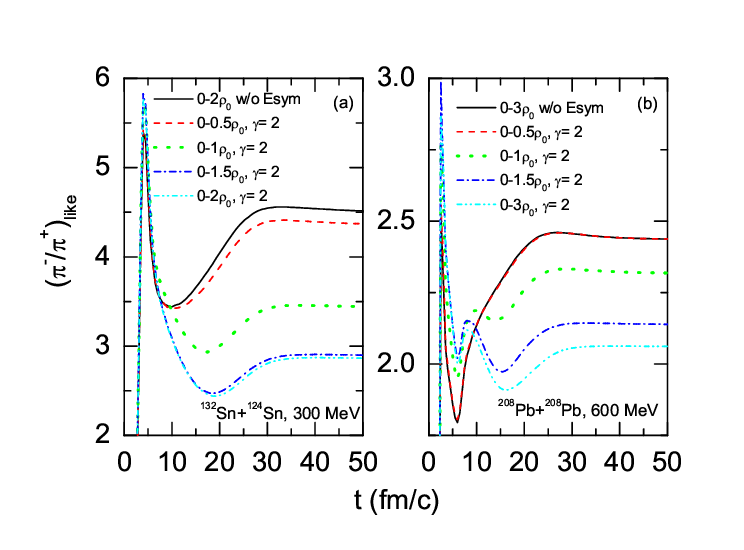}
\caption{\label{rpitime} (Color online) Comparison of the relative sensitivity of the
symmetry energy sensitive observable $(\pi^-/\pi^+)_{like}$ ratio as a function
of time in the $^{132}$Sn+$^{124}$Sn at 300 MeV/nucleon with an impact parameter b= 3 fm (panel (a)) and in the $^{208}$Pb+$^{208}$Pb at 600 MeV/nucleon with an impact parameter b= 0 fm (panel (b)).}
\end{figure}
It is instructive to show time evolution of the total $(\pi^-/\pi^+)_{like}$ ratio \cite{pion4} in light and heavy reaction systems, respectively.
Taking into account the dynamics of resonance production and decays,
$$
(\pi^-/\pi^+)_{like}\equiv \frac{\pi^-+\Delta^-+\frac{1}{3}\Delta^0}
{\pi^++\Delta^{++}+\frac{1}{3}\Delta^+}.
$$
This ratio naturally becomes the final $\pi^-/\pi^+$ ratio after all resonances have decayed.
From Figure~\ref{rpitime}(a), one sees that
the higher values of the $\pi^-/\pi^+$ ratio than the $(N/Z)^{2}$ of the reaction system ($N,Z$ being the neutron number and proton number of the reaction system) indicate that the $\Delta_{1232}$ isobar model \cite{isobar} does not always work well \cite{isobar2,isobar3}.
It is also seen that, in the $^{132}$Sn+$^{124}$Sn reaction at 300 MeV/nucleon, the effects of the symmetry energy in the density region 0-0.5$\rho_{0}$ are larger than that in the density region 1.5$\rho_{0}$-2$\rho_{0}$. The effects of the symmetry energy below saturation density are comparable to that above saturation density. However, the situation is changed when one moves to Figure~\ref{rpitime}(b), evolution of the $\pi^-/\pi^+$ ratio in the $^{208}$Pb+$^{208}$Pb reaction at 600 MeV/nucleon. From Figure~\ref{rpitime}(b), it is seen that, the effects of the symmetry energy in the density region 0-0.5$\rho_{0}$ are obviously smaller than that in the density region 1.5$\rho_{0}$-3$\rho_{0}$. The effects of the symmetry energy below saturation density are also obviously smaller than that above saturation density.
Figure~\ref{rpitime} clearly shows that, to use the $\pi^-/\pi^+$ ratio as the probe of the high-density symmetry energy, using heavy system and at relatively higher incident beam energies is a preferable way.

It is worth mentioning that, although the more complicated transport model simulation integrally shifts the value of the $\pi^-/\pi^+$ ratio \cite{cy16}, the determination of the density region of the symmetry energy probed by the $\pi^-/\pi^+$ ratio should be roughly the same. This is because the maximum pion freeze-out local baryon densities calculated by the complicated transport model \cite{cy16} are about 1.05$\rho_{0}$ for the $^{132}$Sn+$^{124}$Sn reaction at 300 MeV/nucleon and 1.95$\rho_{0}$ for the $^{208}$Pb+$^{208}$Pb at 600 MeV/nucleon, which are roughly equal to that shown in Figure~\ref{pifree}.

\begin{figure}[t]
\centering
\includegraphics[width=0.5\textwidth]{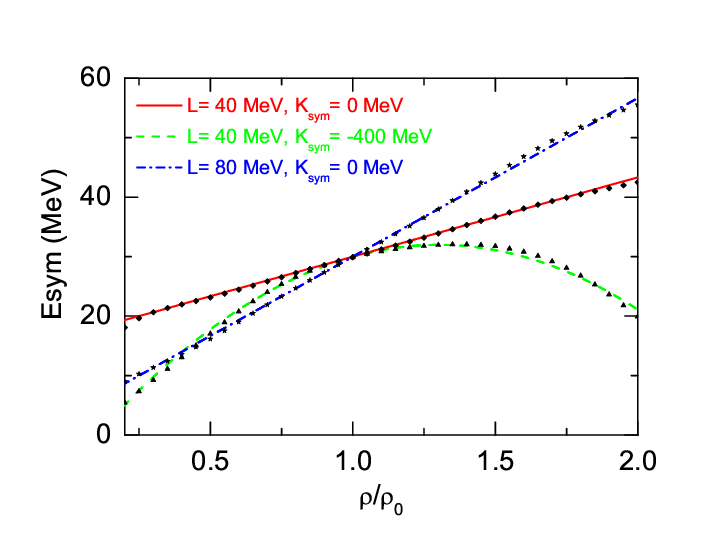}
\caption{\label{esym} (Color online) Density-dependence of the symmetry energy with different slopes and curvatures. The symbols are derived from the momentum-dependent single particle potential used in Ref.~\cite{gyong18}.}
\end{figure}
\begin{figure}[h]
\centering
\includegraphics[width=0.5\textwidth]{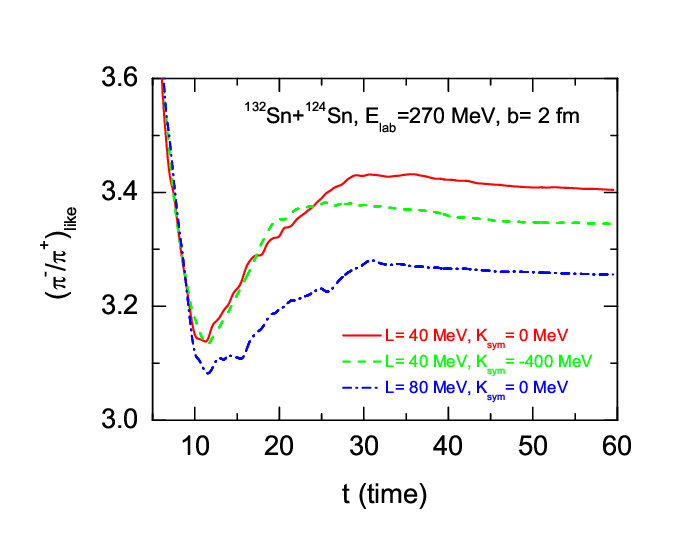}
\caption{\label{lkpi} (Color online) Effects of the slope and curvature of the symmetry energy at saturation density on the $(\pi^-/\pi^+)_{like}$ ratio as a function
of time in the $^{132}$Sn+$^{124}$Sn at 270 MeV/nucleon with an impact parameter b= 2 fm.}
\end{figure}
Similarly, we can also study the density region of the symmetry energy probed by the $\pi^-/\pi^+$ ratio through the Taylor expanded symmetry energy method \cite{gyong18},
$$
E_{sym}(\rho)=E_{sym}(\rho_0)+L\left(\frac{\rho-\rho_0}{3\rho_0}\right)
+\frac{K_{sym}}{2}\left(\frac{\rho-\rho_0}{3\rho_0}\right)^2.
$$
From this expression, it is seen that the slope $L$ is relevant to the symmetry energy around saturation density, but the curvature $K_{sym}$ is more related to the symmetry energy far from saturation point. Whether an observable probes the high-density symmetry energy or not, the simple way is to see if that observable is sensitive to the $K_{sym}$. If the effects of the $K_{sym}$ on the observable is larger than that of the slope $L$, then the observable can be a potential probe of the high-density symmetry energy. In the simulations, different from the RMF effective field \cite{mditoro05,mditoro06}, a Hartree-Fock calculation using the Gogny effective interaction single nucleon potential is used \cite{libanpa04,gyong18}.

Figure~\ref{esym} shows, respectively, the density dependent symmetry energies with the slope $L= 40$ MeV, curvatures $K_{sym}= 0$ and -400 MeV and the case $L= 80$ MeV, $K_{sym}= 0$ MeV. It is seen that the relative magnitudes of slopes of the symmetry energy are different with density.
To see specifically how the slope $L$ and curvature $K_{sym}$ of the symmetry energy at saturation affect the $\pi^-/\pi^+$ ratio, we plot Figure~\ref{lkpi}, effects of the slope and curvature of the symmetry energy at saturation on the $(\pi^-/\pi^+)_{like}$ ratio in the $^{132}$Sn+$^{124}$Sn at 270 MeV/nucleon. It demonstrates that the effect of slope $L$ (changing from 40 to 80 MeV) on the $(\pi^-/\pi^+)_{like}$ ratio is about 3 times larger than the curvature $K_{sym}$ (changing from 0 to -400 MeV). Compared with the soft symmetry energy labeled by $L= 40$ MeV and $K_{sym}= 0$ MeV, the stiff symmetry energy labeled by $L= 80$ MeV and $K_{sym}= 0$ MeV corresponds neutron-deficient matter thus causes a low value of the $\pi^-/\pi^+$ ratio \cite{pion4}. For the symmetry energy labeled by $L= 40$ MeV and $K_{sym}= 0$ MeV, when changing the curvature $K_{sym}$ from 0 to -400 MeV, the somewhat low value of the $\pi^-/\pi^+$ ratio is obtained as shown in Figure~\ref{lkpi}. As shown in Figure~\ref{esym}, the change of the curvature $K_{sym}$ from 0 to -400 MeV stiffens the symmetry energy in the sub-saturation density region, which causes a low value of the $\pi^-/\pi^+$ ratio. That is to say, the observable $\pi^-/\pi^+$ ratio to some extent probes the symmetry energy somewhat below the saturation point. This result is consistent with that shown in Figure~\ref{rpitime} for the $^{132}$Sn+$^{124}$Sn prediction.

\section{Conclusions and discussions}

In summary, to probe the high-density behavior of the symmetry energy, frequently used observable $\pi^-/\pi^+$ ratio is not always suitable. Below 400 MeV/nucleon incident beam energies, for light or medium nucleus-nucleus reaction systems, the $\pi^-/\pi^+$ ratio mainly probes the symmetry energy around saturation density. For the $^{132}$Sn+$^{124}$Sn reactions at 300 or 200 MeV/nucleon, which are being carried out at RIKEN/Japan, the produced $\pi^-/\pi^+$ ratio merely probe the density-dependent symmetry energy around saturation density. To probe the high-density symmetry energy by the $\pi^-/\pi^+$ ratio, a heavy reaction system at relatively high beam energies is preferable.

In the transport model, the in medium thresholds \cite{cozmaplb2016,song2015}, the short range correlations of nucleons, especially the high-momentum tail of nucleon momentum distribution in nuclei \cite{src1,src2,src3,src4}, pion potential and s/p-wave effects \cite{zhangzhen2017,pion7,guoprc2015}, scattering and re-absorption \cite{balid2015,guoprc22015,liq2017,liq20172}, etc., affect pion production in heavy-ion collisions. Since we just study the relative sensitivity of the $\pi^-/\pi^+$ ratio in different density regions to the symmetry energy, all the above uncertain factors to some extend are canceled out. To explore the specific density region that the symmetry energy sensitive observable probed, some other methods are also needed \cite{npa16}. In case the slope and the curvature parameters of the symmetry energy around saturation density are constrained by pion or nucleon observables in the $^{132}$Sn+$^{124}$Sn collisions at 300 or 200 MeV/nucleon incident beam energies, the high-density behavior of the symmetry energy is readily deduced \cite{sym1,sym2}.

\section{Acknowledgments}

This work is supported in part by the National Natural Science
Foundation of China (Grant Nos. 11775275, 11435014, 11475050, 11405128), the 973 Program of China (Grant No. 2013CB834405) and the Zhejiang Province science and technology plan project (2015C33035).

\end{document}